\documentstyle[sprocl]{article}
\input{psfig.sty}
\def\Journal#1#2#3#4{{#1} {\bf #2}, #3 (#4)}

\def\be{\begin{equation}}
\def\ee{\end{equation}}
\def\bea{\begin{eqnarray}}
\def\eea{\end{eqnarray}}

\newcommand{\mb}[1]{\mbox{\boldmath $#1$}} 
\newcommand{\npb}{$\{\mb{k},\mb{\ell},\mb{m},\mb{\bar{m}}\}\;$}

\begin{document}

\title{VACUUM SPACETIMES WITH AN ISOMETRY}
\author{F.~FAYOS}
\address{Departament de F\'{\i}sica Aplicada, UPC, E-08028 Barcelona,
Spain\\E-mail: labfm@ffn.ub.es}
\author{C.~F.~SOPUERTA}
\address{Relativity and Cosmology Group, School of Computer Science and
Mathematics,\\ 
Mercantile House, Hampshire Terrace, PO1 2EG Portsmouth, England\\ 
E-mail: carlos.sopuerta@port.ac.uk}

\maketitle\abstracts{In vacuum space-times the exterior derivative of
a Killing vector field is a two-form that satisfies Maxwell equations 
without electromagnetic sources. Using the algebraic structure of this
two-form we have set up a new formalism for the study of vacuum space-times
with an isometry.}

\section{Introduction}
Symmetries play an important role in physics and in particular in
General Relativity, where they have played an important role in the
search for exact solutions of Einstein's equations.\cite{KSHM,DUSH}
Of particular relevance are Killing symmetries, which are generated
by vector fields satisfying Killing equations: $\xi_{a;b}+\xi_{b;a}=0\,,$
where the semicolon denotes covariant differentiation.
In a given space-time we can look for isometries just by
studying the integrability conditions of the Killing equations,
which are:
\begin{equation}
\pounds(\mb{\xi})\Gamma^a{}_{bc} = \pounds(\mb{\xi})R^a{}_{bcd} =
\pounds(\mb{\xi})R^a{}_{bcd;e_1} = \pounds(\mb{\xi})R^a{}_{bcd;e_1e_2}
= \cdots = 0 \,, \label{uico}
\end{equation}
where $\pounds(\mb{\xi})$ means Lie differentiation along the
Killing vector field (KVF) $\mb{\xi}$,
$\Gamma^a{}_{bc}$ are the Christoffel symbols, and $R^a{}_{bcd}$ the
components of the Riemann tensor. The isometries of a space-time form
a Lie group, which provides a way of classifying space-times in terms
of the Lie group that they admit.  Another important way of classifying
space-times is through the algebraic classification of the Weyl tensor
(the so-called Petrov classification~\cite{KSHM}),
$C_{abcd}$, which, in vacuum, coincides with the Riemann tensor.
The relation between both classifications remains still unclear.
In this communication we describe a new approach to the study of Killing
symmetries in vacuum space-times, based on the introduction of an
algebraic structure associated with the KVF, which allows to establish
connections between the existence of an isometry and the Petrov type of
the space-time.

\section{The algebraic structure associated with an isometry}
An interesting property of KVFs in vacuum space-times, firstly noticed by
Papapetrou,\cite{PAPA} is that their exterior derivative is a 2-form
satisfying Maxwell's equations in the absence of electromagnetic currents.  
This 2-form, which we will call the {\em Papapetrou} field associated with 
$\mb{\xi}$, is given by
\begin{equation}
F_{ab}= \xi_{b;a}-\xi_{a;b} = 2\xi_{b;a} \,. \label{papa}
\end{equation}
In the same way as we endow space-times with the algebraic structure
of the Weyl tensor, we can consider the algebraic structure of
$F_{ab}$ as the algebraic structure associated with the KVF $\mb{\xi}$.
The algebraic classification of a 2-form consists of
two differentiated cases: (i) The {\em regular} case, characterized by
\[ \tilde{F}^{ab}\tilde{F}_{ab}\neq 0\,,  ~~ \mbox{where} ~~
\tilde{F}_{ab}\equiv F_{ab}+i*F_{ab}\,, ~~ \mbox{and} ~~
*F_{ab}\equiv\textstyle{1\over2} \eta_{abcd}F^{cd} \,, \]
and where $\;\tilde{}\;$ and $\ast$ denote the self-dual and dual 
operations respectively.  In this case we can pick a Newman-Penrose basis 
\npb so that $\mb{\tilde{F}}$ takes the following {\em canonical} form
\begin{equation}
\tilde{F}_{ab} = -(\not\!\alpha+i\not\!\beta) W_{ab} \,, ~~ W_{ab}\equiv
2m_{[a}\bar{m}_{b]}-2k_{[a}\ell_{b]} \,, \label{cfre}
\end{equation} 
where \mbox{$\not\!\alpha$} and \mbox{$\not\!\beta$} are the real eigenvalues
of $F_{ab}$, and $\mb{\ell}$ and $\mb{k}$ are its null eigenvectors.
(ii) The {\em singular} case, characterized by
$\tilde{F}^{ab}\tilde{F}_{ab}=0$.  Now, we can choose the Newman-Penrose
basis so that $\mb{\tilde{F}}$ can be cast in the form
\begin{equation}
\tilde{F}_{ab} = \phi V_{ab}\,, ~~ V_{ab}\equiv 2k_{[a}m_{b]}\,,\label{cfsi}
\end{equation}
where $\phi$ is a complex scalar and $\mb{k}$ the only principal direction.
There is a covariant way~\cite{FASO} of obtaining the eigenvalues and
principal direction(s) from the KVF and quantities constructed from it.

Combining the algebraic structures of the Weyl tensor and of the Papapetrou
field of a KVF, we can introduce a new classification of the vacuum 
space-times having at least one KVF, or more precisely, of the pairs 
$\{(V_4,\mb{g}),\mb{\xi}\}$, which takes into account the fact 
that there are space-times with more than one KVF. Then, we can classify
these pairs according to the following properties: The algebraic type of the
Papapetrou field (regular or singular); the Petrov type of the space-time
(I, II, III, D, N, or O); the degree of alignment of the principal directions
of the Papapetrou field with those of the Weyl tensor. Finally, we can 
refine this classification by adding differential invariant properties of 
the principal null directions.

\section{A formalism for vacuum space-times with an isometry}
Starting from the Papapetrou field and its algebraic structure we are going
to introduce a new approach, extension of the Newman-Penrose (NP)
formalism,\cite{NEPE} to study vacuum space-times with a non-null KVF.
This extension consists of two steps: (i) To add new variables related
to the KVF, and their corresponding equations.  (ii)  To write all
the equations (those of the NP formalism plus those for the new variables)
in a NP basis in which the Papapetrou field of the KVF takes its canonical
form [(\ref{cfre}) in regular case and~(\ref{cfsi}) in the singular case].
Then, the alignment of a principal direction of the Papapetrou field with
one of the Weyl tensor can be study in a natural way within this formalism.
For instance, setting $\Psi_0=0$ we impose the principal direction of the
Papapetrou field $\mb{k}$ to be aligned with one principal direction
of the space-time.

The new variables will be, first, the components of the KVF in a NP
basis:
\[ \mb{\xi} = -\xi_l\mb{k}-\xi_k\mb{\ell}+\xi_{\bar{m}}\mb{m}+\xi_m
\mb{\bar{m}}\,, \]
where $\xi_l$ and $\xi_k$ are real and $\xi_m$ complex ($\bar{\xi}_m
=\xi_{\bar{m}}$).  The equations for these variables come from the
definition of the Papapetrou field in terms of the KVF [Eq.~(\ref{papa})],
that is, $\xi_{b;a}=\textstyle{1\over2}F_{ab}\,.$ When we write these
equations in a canonical NP basis~\cite{FSTS}
[See Eqs.~(\ref{cfre},\ref{cfsi})] we realize that the quantities
$(\mbox{$\not\!\alpha$},\mbox{$\not\!\beta$})$ or $\phi$ appear. Then,
in order to close the system of equations we will consider them as new
variables.  It turns out that the equations for these variables are
the Maxwell equations for the Papapetrou field, i.e., $F^{ab}{}_{;b}=0\,,$
$F_{[ab;c]}=0$, which indeed close the system of equations.\cite{FSTS}
In the regular case they are:
\begin{equation}
D(\not\!\alpha+i\not\!\beta)=2\rho(\not\!\alpha+i\not\!\beta)\,, ~~
\triangle(\not\!\alpha+i\not\!\beta)=-2\mu(\not\!\alpha+i\not\!\beta)\,,
\label{max1}
\end{equation}
\begin{equation}
\delta(\not\!\alpha+i\not\!\beta)=2\tau(\not\!\alpha+i\not\!\beta)\,, ~~
\bar{\delta}(\not\!\alpha+i\not\!\beta)=-2\pi(\not\!\alpha+i\not\!\beta)
\,. \label{max2}
\end{equation}
Once the formalism has been set up, let us see how it works.  To that
end, we will focus on the regular case ($\mbox{$\not\!\alpha+i\not\!\beta$}
\neq 0$), since the singular case has already been completely
examined~\cite{FSTS} and all the spacetimes and the KVFs determined:
They correspond to two particular classes of {\em pp waves} (Petrov type N
solutions) and the Minkowski space-time.

One of the things we should look at is the integrability of the
equations~(\ref{papa}) for the components of the KVF, $(\xi_k,\xi_l,\xi_m)$.
The calculations are long but straightforward.  The result is that they
are integrable if the components of the Weyl tensor are given by
the following expressions
\begin{equation}
\Psi_0 = \frac{\not\!\alpha+i\not\!\beta}{N}(\kappa\xi_m -\sigma\xi_k) \,,
~~ \Psi_1 = \frac{\not\!\alpha+i\not\!\beta}{N}(\kappa\xi_l -
\sigma\bar{\xi}_m)\,, \label{exp1}
\end{equation}
\begin{equation}
\Psi_2 = \frac{\not\!\alpha+i\not\!\beta}{N}(\rho\xi_l-\tau\bar{\xi}_m) \,,
\label{exp2}
\end{equation}
\begin{equation}
\Psi_3 =  \frac{\not\!\alpha+i\not\!\beta}{N}(\mu\bar{\xi}_m -
\pi\xi_l) \,, ~~
\Psi_4 = \frac{\not\!\alpha+i\not\!\beta}{N}(\nu\bar{\xi}_m -
\lambda\xi_l) \,, \label{exp3}
\end{equation} 
and the following relations between spin coefficients and components of the
KVF hold
\[ \kappa\xi_l-\sigma\bar{\xi}_m = \rho\xi_m-\tau\xi_k \,, ~~
   \rho\xi_l-\tau\bar{\xi}_m = \mu\xi_k-\pi\xi_m\,, ~~
   \mu\bar{\xi}_m-\pi\xi_l = \nu\xi_k-\lambda\xi_m \,.  \]
As we can see, the procedure used in this formalism determines completely
the components of the Weyl tensor in terms of spin coefficients and
quantities constructed from the KVF, and the dependence on these quantities
is algebraic.  This contrasts with the usual approach to the integrability
conditions for the equations of a KVF which, as we have mention above,
involve derivatives of the curvature [see Eq.~(\ref{uico})].  The main
features of our procedure that are responsible of this result are: The
addition of new variables corresponding to components of the Papapetrou
field, the choice of a NP basis in which the Papapetrou field takes
its canonical form, and finally, we are using the NP formalism which
takes advantage of the fact that the spacetime is a four-dimensional
manifold.  It can be seen that in higher dimensions the expressions
we get for the components of the Weyl tensor do not determine it completely.
A direct consequence of the expressions~(\ref{exp1}-\ref{exp3}) for the
Weyl tensor is that we do not need to solve the second Bianchi identities,
which are the equations for the complex scalars $\Psi_A$. Instead, we only
have to substitute the expressions we have just obtained in the second
Bianchi identities to obtain a set of consistency relations.

Taking into account these results, the next step in the application of
our formalism will be to study the integrability conditions for the Maxwell
equations~(\ref{max1},\ref{max2}), or equivalently,
the integrability conditions for the complex quantity 
\mbox{$\not\!\alpha+i\not\!\beta$}.
Since we know explicitly all the directional derivatives of
\mbox{$\not\!\alpha+i\not\!\beta$}, it is straightforward to study their 
integrability conditions.   
We have found~\cite{FSTS} that they
are additional conditions on the spin coefficients and their directional
derivatives.  Another step forward in this development is to
substitute the expressions~(\ref{exp1}-\ref{exp3}) for the complex
scalars $\Psi_A$ into the second Bianchi identities.  Using the other
equations in our formalism we get expressions containing directional
derivatives only of the spin coefficients, that is, we get more
restrictions on the spin coefficients.\cite{FSTS}
The conclusion we extract from this discussion is that all the integrability
and consistency conditions can be reduce to first-order differential 
equations for the spin coefficients. Specifically they are: (i) The NP 
equations. (ii) The integrability conditions for the Maxwell
equations~(\ref{max1},\ref{max2}).  (iii) The consistency conditions
coming from the substitution of the expressions~(\ref{exp1}-\ref{exp3})
into the second Bianchi identities.

As is clear, the next step in this study would be to look for the
compatibility conditions for the whole set of differential
equations for the spin coefficients.  In general, the calculations
involved are large, but it is possible to study particular cases,
specially those in which there are alignments of the principal direction(s)
of the Papapetrou field with those of the space-time.
For instance, we have studied~\cite{FSTS} Petrov type III vacuum space-times
arriving at the conclusion that the alignment of the multiple principal
direction of the space-time with some of the two principal directions
of the Papapetrou field is forbidden.   In contrast to this situation,
there are other vacuum space-times in which we can find alignments.
An interesting example is the case of the Kerr metric in which the two
multiple principal directions of the space-time (it is Petrov type D) are
aligned with those of the Papapetrou field.\cite{FASO}
A systematic study of alignments in the class of vacuum spacetimes
for which \mbox{$\not\!\alpha+i\not\!\beta$} is an analytic function of the
Ernst potential has been carried out recently.~\cite{FSCO}

To sum up, we have set up a new formalism for the study of vacuum spacetimes
with an isometry.   Taking into account the features we have
described above, this formalism is suitable for the study of any problem or
situation with a Killing symmetry and in which the knowledge of the
Weyl complex scalars is required.  Some interesting examples are:
Search and study
of exact solutions, perturbations of black holes preserving a symmetry, the
question of the equivalence of metrics, the construction of numerical 
algorithms, etc.

\section*{Acknowledgments}
F.F. acknowledges financial support from the D.G.R. of the 
Generalitat de Catalunya (grant 1998GSR00015), and the Spanish Ministry of 
Education (contract PB96-0384). C.F.S. is supported by the European 
Commission (contract HPMF-CT-1999-00149). 


\end{document}